\pdfoutput=1

\documentclass[11pt]{article}

\usepackage{ACL2023}

\usepackage{times}
\usepackage{latexsym}

\usepackage[T1]{fontenc}

\usepackage[utf8]{inputenc}

\usepackage{microtype}

\usepackage{inconsolata}

\usepackage{amsmath,amsfonts,bm}
\usepackage{graphicx}
\usepackage{multirow}
\usepackage{makecell}
\usepackage{array}
\usepackage[para]{threeparttable}
\usepackage{booktabs}
\usepackage{tabularx}
\usepackage{xspace}
\usepackage{caption}
\usepackage{subcaption}
\usepackage{longtable}
\usepackage{bbm}
\usepackage{bm}
\usepackage{enumitem}
\usepackage{algorithm}
\usepackage{algpseudocode}
\usepackage{geometry}
\usepackage{comment}
\usepackage[colorinlistoftodos]{todonotes}
\usepackage{tablefootnote}
\usepackage{color}

\newcommand{\mat}{CE\xspace}
\newcommand{\Npd}{Nonparametric Decoding\xspace}
\newcommand{\npd}{Np Decoding\xspace}

\newcommand{\Pipe}{\textsc{Contra} Np Decoding\xspace}
\newcommand{\Base}{\textsc{Base} Np Decoding\xspace}

\newcommand{\async}{\textsc{Async}\xspace}
\newcommand{\pipe}{\textsc{Contra}\xspace}
\newcommand{\base}{\textsc{Base}\xspace}
\newcommand{\basen}{\textsc{Base}-Short\xspace}
\newcommand{\asyncn}{\textsc{Async}-Short\xspace}
\newcommand{\emb}{CE Encoder\xspace}
\newcommand{\ret}{generative retriever\xspace}

\title{Nonparametric Decoding for Generative Retrieval}

\author{Hyunji Lee$^{1}$ \quad Jaeyoung Kim{\textsuperscript{3}\thanks{\, Work done during internship at KAIST AI.}} \quad Hoyeon Chang$^{1}$ \quad Hanseok Oh$^{1}$ \quad Sohee Yang$^{1}$\\
{\bf Vlad Karpukhin$^{2}$} \quad {\bf Yi Lu$^{2}$} \quad {\bf Minjoon Seo$^{1}$} \\
 \\
$^{1}$ KAIST AI \hspace{3.0em} $^{2}$ Forethought.AI \hspace{3.0em} $^{3}$ Kakao\vspace{0.5em} \\
\texttt{\{hyunji.amy.lee, hanseok, sohee.yang, retapurayo, minjoon\}@kaist.ac.kr} \\
\texttt{\{vlad.karpukhin, yi.lu\}@forethought.ai}\hspace{3.0em} \texttt{jay.eong@kakaocorp.com}
  }

\begin{document}
\maketitle
\begin{abstract}

The generative retrieval model depends solely on the information encoded in its model parameters without external memory, its information capacity is limited and fixed. 
To overcome the limitation, we propose \Npd~(\npd) which can be applied to existing generative retrieval models.
\npd uses nonparametric contextualized vocab embeddings (external memory) rather than vanilla vocab embeddings as decoder vocab embeddings.
By leveraging the contextualized vocab embeddings, the generative retrieval model is able to utilize both the parametric and nonparametric space.
Evaluation over 9 datasets (8 single-hop and 1 multi-hop) in the document retrieval task shows that applying \npd to generative retrieval models significantly improves the performance.
We also show that \npd is data- and parameter-efficient, and shows high performance in the zero-shot setting.\footnote{The code and datasets used in our work is at \\ \href{https://github.com/amy-hyunji/Contextualized-Generative-Retrieval}{https://github.com/amy-hyunji/Contextualized-Generative-Retrieval}.}
\end{abstract}

\section{Introduction}

Text retrieval is often formulated as finding the most relevant items from a large corpus given an input query. 
The bi-encoder approach of using an encoder to map the documents and the query to a common vector space and performing a nearest neighbor search has been a common practice in text retrieval tasks~\citep{karpukhin2020dense, Wu2020ScalableZE, ni2021sentence}. 
Despite its high performance and popularity, it has an embedding space bottleneck~\citep{luan2021sparse, Lee2022GenerativeMR}; limited expressiveness due to fixed-size embeddings and misses the fine-grained interaction between embeddings as they interact in L2 or inner product space. 
Moreover, the bi-encoder approach requires large storage space to save all document embeddings.

A recently-proposed alternative to the bi-encoder approach is using a generative retrieval model~\citep{cao2021autoregressive, Tay2022TransformerMA, Bevilacqua2022, Lee2022GenerativeMR, Wang2022NCI, Lafferty2003ProbabilisticRM, Croft2010LanguageMF}. It is an autoregressive model that retrieves the most relevant sequence by generating the target sequence (e.g., title, passage, document ID) token-by-token. 
It overcomes the embedding space bottleneck by interacting in the parametric space. Also, it is storage efficient by not having any external memory.
However, the information capacity of such fully parametric models tends to be bounded by their sizes as it has to encode all information in its parameters~\citep{Tay2022TransformerMA, Roberts2020HowMK}. 

\begin{figure*}
\centering
\includegraphics[width=0.8\textwidth]{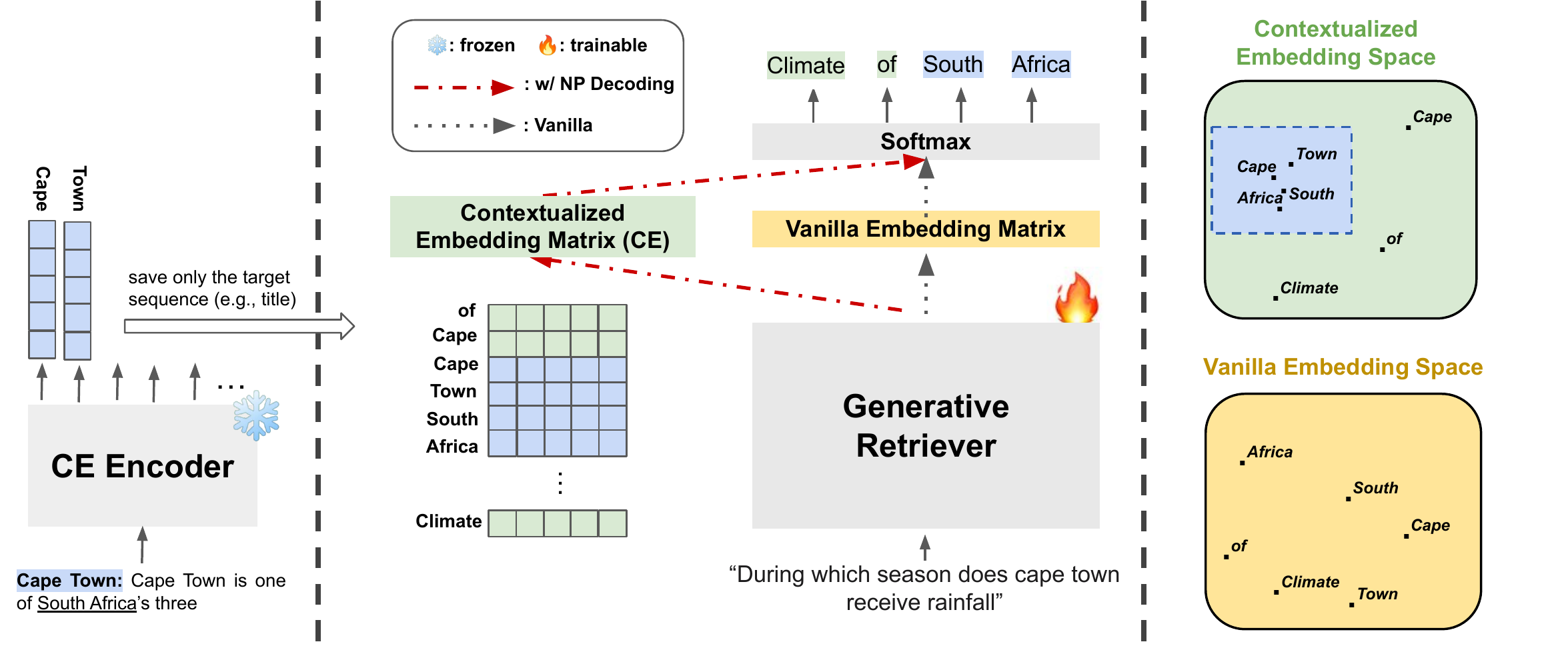}
\caption{\fontsize{7.5}{10}\footnotesize 
\npd can be applied to any generative retrieval model by replacing the decoder vocab embeddings from the vanilla embedding matrix with the contextualized embedding matrix~(\mat). \mat is composed of the output embeddings of the language model encoder (\emb). Only the retrieval target sequences are added to \mat, which in this figure we use the title (Cape Town) as the target sequence. Unlike vanilla vocab embeddings, contextualized vocab embeddings that consist \mat contain context information, and a single token can have multiple token embeddings. This creates a more expressive and fine-grained contextualized embedding space compared to vanilla embedding space as shown on the right side of the figure.}
\label{fig:base c-aer}
\vspace{-0.5em}
\end{figure*}

To this end, we propose \Npd~(\npd), a decoding method for generative retrieval models. It uses nonparametric contextualized vocab embeddings rather than vanilla vocab embeddings as decoder vocab embeddings. 
The contextualized vocab embeddings are output embeddings of an encoder that constructs a nonparametric dense vector space and are frozen during the training step whereas the vanilla vocab embeddings are trainable model vocab embeddings that construct a parametric space of the model. 
Therefore, by using \npd, the generative retrieval model does not have to rely solely on its own parameters but can utilize the surrounding information encoded in the contextualized vocab embeddings (external memory).
Note that while it utilizes the dense vector space as in the bi-encoder approach, unlike the approach, it does not have embedding space bottleneck as it is a variant of the generative retrieval model, and saves storage space by storing only clustering centroid embeddings (Section~\ref{sec4: cluster}).

As shown in Figure~\ref{fig:base c-aer}, any generative retrieval model can incorporate \npd by replacing the decoder vocab embeddings from the vanilla embedding matrix to contextualized embedding matrix (\mat) for both the training and the inference steps.
By the replacement, \npd has two key benefits over vanilla decoding.
First, the generative retrieval model can utilize not only its parametric space but also its nonparametric space. The nonparametric space is constructed with decoder vocab embeddings of \npd (\mat), nonparametric and context-aware embeddings that capture surrounding information.
Second, \mat allows a token to have multiple token embeddings, unlike vanilla vocab embeddings where a token has a unique embedding. Therefore, the decoder vocab embedding space of \mat becomes more expressive and fine-grained (right side of Figure~\ref{fig:base c-aer}).
Since having a well-constructed \mat is important for achieving high performance, we propose three different encoders (\emb) used to output contextualized vocab embeddings added to \mat (Section~\ref{Sec3}). 
We demonstrate that \emb with contrastive learning results in a significant increase in performance.  

The main contributions of our paper are as follows:
\begin{itemize}[noitemsep,leftmargin=1.8em]
    \item We propose \Npd~(\npd), a simple and novel decoding method that can be applied to all existing generative retrieval models. Experimental results over 9 datasets show that \npd can significantly improve the performance of existing generative retrieval models by leveraging both the parametric and the nonparametric space; 4.4\% R-precision improvement for single-hop, 5.4\% Recall@2 improvement for multi-hop datasets.
    \item We present various \emb and show that training \emb with contrastive learning further increases the performance by a large margin.
    \item We show generative retrieval models with \npd are data- and parameter-efficient, and show higher performance in a zero-shot setting.
\end{itemize}

\section{Related Work}

\paragraph{Generative Retrieval} \label{related: gr}
Generative retrieval models retrieve relevant items by generating sub/either the identifiers or entire sequences of the items. 
GENRE~\citep{cao2021autoregressive} retrieves a document by generating the titles with a constrained beam search. 
DSI~\citep{Tay2022TransformerMA} assigns a unique ID to each item in the corpus and retrieves the item by generating the ID of the most relevant document.
SEAL~\citep{Bevilacqua2022} retrieves any span from any position in the corpus by using FM-Index. 
GMR~\citep{Lee2022GenerativeMR} retrieves the most relevant item by generating the whole sequence.
Though high performance, as generative retrieval models solely rely on the information stored in their parameter, the information capacity is limited and fixed. 
To overcome the limitation, we propose \Npd~(\npd) for generative retrieval models. By replacing the decoder vocab embeddings with nonparametric contextualized vocab embeddings, the model is able to utilize not only the parametric space but also the nonparametric space of contextualized embeddings.  

\paragraph{Memory Augmented Models}
KNN-LM~\citep{Khandelwal2020GeneralizationTM}, TRIME~\citep{Zhong2022TrainingLM}, RAG~\citep{lewisretrieval}, and RETRO~\citep{Borgeaud2022ImprovingLM} are memory augmented models which use both the parametric space of the model and the non-parametric space of the external memory.
KNN-LM improves the LM performance by generating the next token through interpolation between the nearest neighbor distribution (distance in the contextualized embedding space) and the model vocab distribution only during the inference step. TRIME expands the work to use the objective also during the training step.
RAG and RETRO first retrieve relevant texts with the retriever from the external memory and generate the output based on the retrieved texts.
Moreover, concurrent work NPM~\citep{Min2022NonparametricML} proposes a nonparametric masked language model which operates over the nonparametric distribution of the external memory.
Generative retrieval models with \Npd also utilize the external memory, but rather than considering it as an external source, it is incorporated with the model by utilizing the external memory as decoder vocab embeddings.

\section{\Npd} \label{Sec3}
Generative retrieval is the task of retrieving the most relevant retrieval target (e.g., title, passage, document identifier) by generating the target token-by-token when given an input query.
The training objective of the generative retrieval model is to maximize 
\begin{equation} \label{eq: gr}
P((t_1, \cdots, t_n)|q) = \prod_{i=1}^{n} P(t_i|q, t_{<i}) 
\end{equation}
where $t_*$ denotes the tokens of the retrieval target and $q$ is the input query.
Such an approach has shown high performance while using a low storage footprint~\citep{cao2021autoregressive, Tay2022TransformerMA, Bevilacqua2022, Lee2022GenerativeMR}.
However, it has limitation in that the model depends \textit{solely} on the information encoded in its own parameters. 
Thus, the performance is likely to be bounded by how much information can be stored in the model parameter~\citep{Tay2022TransformerMA, Roberts2020HowMK}.

To address the limitation, we propose a new decoding method called \Npd~(\npd) for generative retrieval. 
To incorporate \npd on the existing generative retrieval model, the only amendment is to use the frozen contextualized vocab embedding (external memory) rather than the vanilla vocab embedding as the decoder vocab embedding during each generation step (Figure~\ref{fig:base c-aer}). 
The embeddings are the output embeddings of an encoder when given a target sequence as input. 
Note that existing generative retrieval models such as GENRE and DSI utilize the pre-trained language model architecture as-is: vanilla vocab embedding as the decoder vocab embedding.

In Section~\ref{sec3: benefits}, we show the key benefits of using \npd over vanilla decoding.
For Section~\ref{sec3:base-cgr} to Section~\ref{sec3:pipeline-cgr}, we show the details of base \npd (\base), and two variants (\async, \pipe).
In Section~\ref{sec4: cluster}, we describe how we reduce the number of contextualized token embeddings.

\subsection{Key Benefits} \label{sec3: benefits}

Using \npd has two key benefits over vanilla decoding.
First, the generative retrieval model with \npd can utilize not only the information encoded in its own parameters (parametric space) but also the surrounding information encoded in the contextualized vocab embeddings (nonparametric space) during each decoding step. 
Second, the generative retrieval model with \npd has more expressive and fine-grained decoder vocab embedding space than that of the model with vanilla decoding. 
As in Figure~\ref{fig:base c-aer}, \npd allows a single token to have multiple contextualized token embeddings for the decoder vocab embeddings (e.g., the same token "Cape" has two different contextualized embeddings) depending on the surrounding information of the token, whereas vanilla decoding allows only a single token embedding for a single token. Note that we do not save all possible token embeddings, but reduce the number of tokens to save without performance degradation by practical tactics (Section~\ref{sec4: cluster}).


\subsection{\textsc{Base} \Npd}\label{sec3:base-cgr}

In this work, we propose three different \npd (Base \Npd and two variants) which we name the three different \npd based on the characteristics of the Contextualized Embedding Encoders~(\emb).
\emb is an encoder that outputs contextualized token embeddings when given a target sequence (e.g., title, document ID, passage) as input. 
The contextualized token embeddings are added to \mat\footnote{Details of how we construct \mat for different target sequences are in Section~\ref{exp: details}.}, the decoder vocab embedding matrix of generative retriever with \npd.
\textsc{Base} \Npd~(\base) uses the most basic \emb, the pre-trained T5 encoder as-is. \mat is constructed once with the output embeddings of \emb before the generative retrieval training step.
Note that during the training step of the generative retrieval, \emb is frozen (Figure~\ref{fig:base c-aer}).

\subsection{\textsc{Async} \Npd} \label{sec3:iter-cgr}

Asynchronous Nonparametric Decoding~(\async) uses \emb which is \textit{asynchronously replaced} every $N$ epoch by the encoder of generative retriever during the generative retrieval training step. By replacing \emb periodically, \async has more coherency between \emb and the generative retriever than \base.
After every replacement ($N$ epoch), we construct a new \mat with the output embeddings of replaced \emb and resume training the generative retriever.
Note that during the generative retrieval training step, \emb is frozen but simply replaced, and only \ret is trainable.
We keep $N=20$ for all experiments. See Appendix~\ref{ablation: freq} for details on how $N$ affects the performance.


\begin{figure}
\centering
\includegraphics[width=0.45\textwidth]{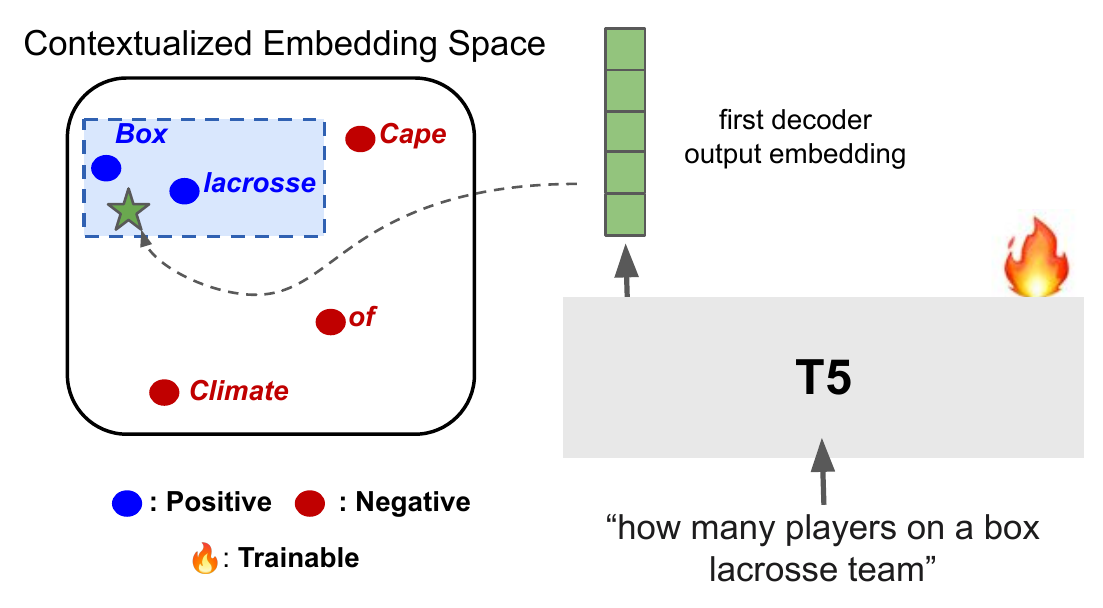}
\caption{\fontsize{7.5}{10}\footnotesize Token-level contrastive learning of \Pipe. Given a query ("how many players on a box lacrosse team") and target sequence ("Box lacrosse"), we train T5 on token-level contrastive learning where all tokens of the target sequence are the positive pairs and the rest of the tokens in \mat are negative pairs.} 
\label{fig:contra}
\vspace{-1.0em}
\end{figure}



\subsection{\textsc{Contrastive} \Npd} \label{sec3:pipeline-cgr}

\textsc{Contrastive} \Npd (\pipe) uses \emb trained on \textit{token-level contrastive learning}. 
The \emb constructs \mat, the nonparametric decoder vocab space of generative retrieval model with \npd. 
The token-level contrastive learning (Equation~\ref{eq: contra}) is performed as an intermediate step before training T5 on the generative retrieval task (Equation~\ref{eq: gr}). 
Bi-encoder retrieval models with contrastive loss have shown high performance as the model learns to construct well-structured global embedding space and regularize the space to be uniform~\citep{ni2021sentence, Gao2021SimCSESC, Gao2022UnsupervisedCA, Izacard2021UnsupervisedDI}. 
In a similar way, \emb with contrastive learning constructs a more meaningful dense vector space (non-parametric space of the generative retriever) than \emb of \base.


As in Figure~\ref{fig:contra}, given a query, we train the first output embedding of the T5 decoder\footnote{We use the embedding of decoder~\citep{ni2021sentence}, not the encoder, to initialize generative retriever with both the encoder and the decoder trained on contrastive learning.} with all tokens of the target sequence as positive pairs and the rest of the tokens in \mat\footnote{As we freeze the token embeddings (\mat) and only train the T5, calculating over entire embedding space is possible. \mat used in the step is constructed with the output embeddings of the pre-trained T5 encoder model.} as negative pairs. 
After training T5 with token-level contrastive learning, we construct the CE with its encoder as \emb, and then further train the model on the generative retrieval task. 

\paragraph{Step 1. Token-level Contrastive Learning}
Given a training dataset of pairs $\{(\textbf{\textit{q}}, \textbf{\textit{t}})\}$ where $\textbf{\textit{q}}$ is the query text, and $\textbf{\textit{t}}$ is the retrieval target (e.g., the title of the document to retrieve) composed of multiple tokens $\textbf{\textit{t}}_{i}$ ($1 \le i \le k$ where $k$ is the length of the target), we split the training dataset into $k$ separate pairs $\{(\textbf{\textit{q}}, \textbf{\textit{t}}_i)\}$ to construct the training dataset of query-token.
With the query-token dataset, we train the first output token embedding from the T5 decoder to be close to all token embeddings in $\mathcal{T}^+$ when given query $\textbf{\textit{q}}$ as an input to \ret (Figure~\ref{fig:contra}).
$\mathcal{T}^+$ is a set of positive token embeddings\footnote{$\mathcal{T}^+ = \{\textbf{t}_1^+, \cdots, \textbf{t}_k^+\}$ ($k = |\mathcal{T}^+|$)} (tokens that make up one retrieval target), and $\mathcal{T}^-$ is the set of negative token embeddings\footnote{$\mathcal{T}^- = \{\mathbf{t}_1^{-}, \cdots, \mathbf{t}_{|\mathcal{T}^-|}^{-}\}$} (all other token embeddings in \mat).
The objective is to minimize the contrastive loss: 
\begin{equation}\label{eq: contra}
\begin{split} 
L(\mathbf{q}, \mathbf{t}_1^{+}, \cdots, \mathbf{t}_{|\mathcal{T}^+|}^{+},\mathbf{t}_1^{-}, \cdots, \mathbf{t}_{|\mathcal{T}^-|}^{-}) \\
= -\log \frac{\sum_{\mathbf{t}^+ \in \mathcal{T}^+}{e^{<\mathbf{q}, \mathbf{t}^+>}}}{\Sigma_{\mathbf{t}^+ \in \mathcal{T}^+}{e^{<\mathbf{q}, \mathbf{t}^+>}} + \Sigma_{\mathbf{t}^- \in \mathcal{T}^-}{e^{<\mathbf{q}, \mathbf{t}^->}}}
\end{split}
\end{equation}
where $\langle  \; , \; \rangle$ is the inner product value between the two embeddings.
We also experiment with a contrastive loss having a single token per target as positive and in-batch negatives loss (Appendix~\ref{app: pipe_details}) where the contrastive loss with multiple tokens (Equation~\ref{eq: contra}) as positive shows the highest performance, which we hypothesize is because the positives have similar content information encoded.

\paragraph {Step 2. Generative Retrieval}
After training T5 with token-level contrastive learning, we use the trained encoder as \emb and construct a new \mat.
We then further train the model on the generative retrieval task using the newly constructed \mat as the decoder vocab embeddings\footnote{The generative retrieval model is initialized with the T5 trained with token-level contrastive learning.}.

\subsection{Clustering} \label{sec4: cluster}
To construct \mat, the decoder vocab embedding matrix of \ret, we first extract all contextualized embeddings of each target token with \emb.
As it requires a large storage footprint to save all the embeddings, we reduce the number of embeddings by using clustering and saving only the representative embeddings of each cluster.
To be specific, we perform k-means clustering over the contextualized embeddings of the same token (which might have different surrounding contexts) and leave only the $k$ centroid embeddings\footnote{When the number of extracted contextualized embeddings of a token is smaller than $k$, we do not perform k-means clustering but use its own contextualized embedding. Also, we use a single non-contextualized embedding for special tokens such as the EOS token or PAD token.} as the decoder vocab embeddings of the token.
We keep $k=5$ for all experiments. When $k=5$, it only requires 0.3\% of storage footprint compared to when saving all contextualized token embedding. Also, it requires only 0.34GB more storage compared to the vanilla vocab embeddings ($k=1$) which is marginal compared to the storage footprint to save the model parameters (3GB).
See Appendix~\ref{app: clustering} for details.

\begin{table*}[t!]
\vspace{-3em}
\begin{minipage}{0.6\linewidth}
\centering
\fontsize{7.5}{10}\selectfont
    \begin{tabular}{c|ccccccc|c}
    \toprule
         Model & FEVER & AY2 & TREX & zsRE & NQ & TQA & WOW & Avg\\
    \midrule
    BM25 & 38.2 & 1.4 & 57.4 & 66.3 & 23.4 & 25.2 & 23.9 & 33.7\\
    DPR & 73.0 & 44.6 & 72.9 & 94.1 & \textbf{60.1} & 63.9 & 36.5 & 63.6 \\
    \midrule
    \underline{GENRE*} & 70.3 & \textbf{75.6} & 73.9 & 97.0 & 51.8 & 65.0 & 59.2 & 70.4\\
    G*-\base & 73.3 & 73.0 & \underline{79.2} & \underline{99.1} &59.0 & 68.2 & 61.1 &73.3\\
    G*-\async& \underline{74.7} & 74.6 & 78.2 & 98.9 & 59.2 & \underline{68.4} & \underline{61.9}& \underline{73.7}\\
    G*-\pipe& \textbf{77.1} & \underline{75.5} & \textbf{81.3} & \textbf{99.2} & \underline{59.8} & \textbf{68.6} & \textbf{62.4}&\textbf{74.8}\\
    \bottomrule
    \end{tabular}
\caption
     {\fontsize{6.5}{10}\footnotesize R-precision(\%) for document retrieval task on test dataset when trained with each dataset (KILT version). G*-\base, G*-\async, and G*-\pipe are the results when adding \npd on GENRE*. The best and second best of each dataset in \textbf{bold} and \underline{underline} respectively.}
\label{table: kilt}
\end{minipage}
\hspace{1.0em}
\begin{minipage}{0.4\linewidth}

\centering
\fontsize{7.5}{10}\selectfont
    \begin{tabular}{c|cc}
    \toprule
    & Hits@1 & Hits@10 \\
    \midrule
    BM25 & 11.6 & 34.4 \\
    Sentence-T5 & 22.4 & 63.3 \\
    \midrule 
    DSI$_{\text{Naive}}$ & 13.3 & 33.6 \\
    DSI$_{\text{Semantic}}$ & 35.6 & 62.6 \\
    DSI$_{\text{Naive}}$-\base& 58.7&73.1\\
    DSI$_{\text{Naive}}$-\pipe & \textbf{60.4}&\textbf{75.8}\\
    \midrule
    GENRE* & 53.7 & 64.7 \\
    GENRE*-\base & 62.2&78.8\\
    GENRE*-\pipe & \textbf{63.4}&\textbf{81.1}\\
    \bottomrule
    \end{tabular}
\caption
     {\fontsize{7.5}{10}\footnotesize Hits@1 and @10 in NQ-320k. Results of BM25, Sentence-T5, DSI$_\text{Naive}$, and DSI$_\text{Semantic}$ are from \citet{Tay2022TransformerMA}. Best of each section in \textbf{bold}.}
\label{table: NQ320k}
\end{minipage}
\vspace{-1em}
\end{table*}

\section{Experimental Setup}
In section~\ref{exp: baselines} and section~\ref{exp: dataset}, we describe the baselines and the datasets we used for experiments.
In section~\ref{exp: details}, we show how we construct \mat depending on which \ret we combined with.
We experiment over 9 datasets.
Results show that by simply replacing the decoding strategy, generative retrieval shows significantly higher performance.
See Appendix~\ref{app:exp_ret} for more details about setups.

\subsection{Baselines} \label{exp: baselines}
BM25~\citep{BM25} is a term-matching model relying on an efficient algorithm.
DPR~\citep{karpukhin2020dense} is a bi-encoder retrieval model which retrieves the most relevant document by performing a nearest neighbor search over dense vector space.  
Sentence-T5~\citep{ni2021sentence} is similar to that of DPR but with T5~\citep{raffel2020exploring} as the base model.
MDR~\citep{xiong2021answering} is an extension of DPR to multi-hop datasets by iterating over a single query. More details about the baselines are in Appendix~\ref{app: setup2}.
See Section~\ref{related: gr} for descriptions of GENRE~\citep{cao2021autoregressive}, DSI~\citep{Tay2022TransformerMA}, and GMR~\citep{Lee2022GenerativeMR}. 

\subsection{Datasets \& Evaluation Metrics} \label{exp: dataset}
We use 9 datasets with various characteristics: FEVER~\citep{Thorne2018FEVERAL}, AY2~\citep{Hoffart2011RobustDO}, TREX~\citep{ElSahar2018TRExAL}, zsRE~\citep{Levy2017ZeroShotRE}, NQ~\citep{Kwiatkowski2019NaturalQA}, TQA~\citep{Joshi2017TriviaQAAL}, WOW~\citep{Dinan2018WizardOW}, NQ-320k~\citep{Tay2022TransformerMA}, and HotpotQA~\citep{Yang2018HotpotQAAD}.
For all datasets except for NQ-320k and HotpotQA, we use dataset and corpus from KILT~\citep{petroni-etal-2021-kilt}. 
To compare with DSI~\citep{Tay2022TransformerMA}, we experiment over NQ-320k, a restricted setting from the official NQ dataset; it uses about 4\% of Wikipedia as the corpus set. Note that the NQ of the KILT version and the official version is different (Details are in \citet{petroni-etal-2021-kilt}). HotpotQA~\citep{Yang2018HotpotQAAD} is an open-domain multi-hop question-answering dataset that needs two Wikipedia pages to answer the question. For HotpotQA we used the official version of the dataset and the corpus.

We evaluate all results of the KILT version with R-precision, a metric widely used to evaluate retrieval performance in KILT. It is calculated as $\frac{r}{R}$ where $R$ is the number of Wikipedia documents in each provenance set, and $r$ is the number of related documents among the top-$R$ retrieved documents.
The results of NQ-320k and HotpotQA are evaluated using Hits@N (N=\{1, 10\}), which shows the proportion of the correct documents ranked in the top N predictions.

\subsection{Details of Constructing \mat} \label{exp: details}
The choice of which output embeddings of \emb to save when constructing \mat depends on the target sequence of the generative retrieval model. In this work, we focused on applying \npd to generative retrieval models that use representative words as the target sequence (DSI and GENRE) and leave extending the work to generative retrieval models that need the whole sequence as the target sequence (GMR and SEAL) as future work.

\textbf{\npd on GENRE*}
As the target sequence of GENRE*\footnote{GENRE* is GENRE trained with T5-large. We used T5 as a base model for a fair comparison with other base models. See Appendix~\ref{app: kilt-multi} for a performance comparison between GENRE* (T5-large) and GENRE (Bart-large).} is the title of the most relevant document, we construct \mat with the output embeddings of document titles. 
To additionally encode the content of the document in the title embeddings, we input the title and the document content\footnote{We use the first five paragraphs of the document to the maximum of 512 tokens as the document content due to limited input length.} into \emb and save only the output embeddings of the title when constructing \mat. 

\textbf{\npd on DSI}
As the target sequence of DSI is the document ID of the most relevant document, we construct \mat with the document ID\footnote{As the document ID of official DSI is not released, we assign the document ID with arbitrary unique integers (naively structured identifiers, DSI$_\text{Naive}$). DSI$_\text{Naive}$ with our document ID shows 12.5 and 22.4 for Hits@1 and Hits@10.}. 
As in GENRE*-\npd, we input the document ID with the document content as the input of \emb.

\section{Experimental Results} \label{exp: results}
In this section, we demonstrate the benefits of using \npd in generative retrieval models by comparing the performance of existing generative retrieval models (DSI, GENRE*) with and without the method in document retrieval tasks. 

\subsection{Vanilla Decoding vs. \npd} \label{ret: single}
Table~\ref{table: kilt} shows that using \npd improves task-specific performance on various datasets, with an average of 4.4\% improvement. 
The improvement is especially significant on datasets with a large number of training examples (FEVER, TREX), which we assume is because the model can learn more about the new vocab embeddings (\mat) during the training step.

Table~\ref{table: NQ320k} shows the results of NQ-320k when applying \npd on DSI and GENRE*.
The two models differ in that the retrieval target of DSI is document ID and that of GENRE* is document title.
\npd enhances DSI and GENRE* by 24.8\% and 9.7\% in Hits@1.
The results suggest that applying \npd is especially helpful for cases where the vanilla vocab embeddings of the target sequence have less information; improvement is higher in DSI than in GENRE*. 
As the document ID is constructed with arbitrary unique integers, the vanilla vocab embeddings of document ID contain less semantic information and have not seen the relationship between the document and the ID during the pretraining step.
In contrast, as the title uses vanilla vocab embeddings of natural language, the embeddings would contain more information compared to that of the document ID. 

Table~\ref{table: Hotpot} shows that using \npd also improves performance on HotpotQA, the multi-hop retrieval dataset; GENRE*-\pipe shows a 7\%-increase in performance compared to GENRE* in Recall@2. 
Also, compared to GMR, as GENRE*-\pipe is able to capture the entire context information with just title generation by using the contextualized embeddings, it shows higher performance and faster inference speed.
Furthermore, when compared to a multi-hop bi-encoder model MDR-, a variant of MDR without advanced techniques like linked negatives, memory bank, and shared encoder, GENRE*-\pipe shows higher performance, whereas lower performance compared to MDR, a model with all the techniques are applied. We expect that applying such techniques to generative retrieval models would also be helpful and leave it as a future work.
More details about the results and how we extend GENRE* in the multi-hop setting are in Appendix~\ref{app: hotpot}.

\subsection{Benefits of \Npd} \label{ret: effect}
We found three major benefits of using \npd over vanilla decoding for generative retrieval.

\paragraph{(1) Parameter-Efficient}
GENRE*-\base trained with T5-base (54.0 and 66.4) shows higher performance in NQ and TQA compared to GENRE* trained with T5-large (51.8 and 65.0) while T5-large has 3.5 times more parameters than T5-base. 
Also, DSI-\base (58.7) shows about $1.5 \times$ higher performance in NQ-320k Hits@1 compared to DSI T5-XXL with Semantic String Docid (the best-performing model in \citet{Tay2022TransformerMA}) (40.4), which has 14 times more parameters than DSI-\base, demonstrating that a retrieval model with \npd is more parameter efficient.

\paragraph{(2) Data-Efficient}
GENRE*-\pipe trained on NQ and TQA together has similar performance to GENRE trained on the entire KILT dataset together, despite having only 5\% as many training examples.
To be specific, when evaluated on R-precision, GENRE*-\pipe performs 60.3/68.9, and GENRE performs 60.3/69.2 in NQ/TQA, respectively (Table~\ref{table: kilt_multi} in Appendix~\ref{app: kilt-multi}).
Such results demonstrate that using \npd is advantageous in the low-resource setting as it can utilize the information in the non-parametric space.

\paragraph{(3) Robust to Zero-Shot}
Table~\ref{table: zeroshot} shows that GENRE*-\base is stronger than GENRE* in the KILT zero-shot setting where both models are trained on NQ and TQA together and are evaluated on the other 9 datasets in KILT that are not used during the training step. GENRE*-\base shows an average of 3\% improvement over GENRE*. 
GENRE*-\base is able to generalize well to out of domains as it does not rely only on the information encoded in the parametric space but also utilizes the nonparametric space of \mat which is shared across all domains; \npd makes the generative retrieval model more robust in the zero-shot setting.

\paragraph{(4) Robust to Low Lexical Overlap}
To evaluate whether the model leverages the information encoded in \mat when using \npd, we test the performance of GENRE*-\base and GENRE* on queries that are likely to require utilizing information from document content (queries with low lexical overlap with the target sequence) in order to find the answer. 
We divide the queries in the NQ dev set into low- and high-overlap sets using the TF-IDF score. GENRE* and GENRE*-\base both show relatively high performance on queries in the high-overlap set compared to the low-overlap set as it is easier to infer the correct retrieval target from the query alone even if the model does not know the document content. 
However, GENRE*-\base shows about 7\% higher performance on the low-overlap set and 5\% higher performance on the high-overlap than GENRE*. This shows that GENRE* with \npd (GENRE*-\base) is robust on queries in the low-overlap set by utilizing the information encoded in \mat.
(More details in Appendix~\ref{app:lexical}.)

\subsection{What is Well-Constructed Contextualized Embedding Matrix (\mat)?} \label{ret: mat}
We found that the choice of \mat plays an important role in the performance when applying \npd. We analyzed four factors that are especially important to create a well-constructed contextualized embedding matrix~(\mat). See Appendix~\ref{app: ce} for various analyses of contextualized token embeddings and more details of each subsection.

\begin{table}[t!]
\centering
\fontsize{7.5}{10}\selectfont
\caption
     {\fontsize{6.5}{10}\footnotesize Recall rate of HotpotQA official full-wiki dev set. Results of DPR, MDR-, and MDR are from \citet{xiong2021answering} and results of GMR are from \citet{Lee2022GenerativeMR}. MDR- indicates a variant of MDR without linked negatives, memory bank, and shared encoder. Best among each method in \textbf{bold}.}
    \begin{tabular}{c|ccc}
    \toprule
     Method & Model& Recall@2 & Recall@10\\
    \midrule
    \multirow{3}{*}{Bi-Encoder}
    &DPR & 25.2 & 45.4 \\
    &MDR- & 59.9 & 70.6 \\
    &MDR & \textbf{65.9} & \textbf{77.5} \\
    \midrule
    \multirow{4}{*}{Generative}
    &GMR & 57.7 & 58.8 \\
    &GENRE* & 56.1 & 58.4 \\
    &GENRE*-\base & 61.9 & 65.3  \\
    &GENRE*-\pipe & \textbf{63.1} & \textbf{66.8}  \\
    \bottomrule
    \end{tabular}
\label{table: Hotpot}
\vspace{-1em}
\end{table}

\begin{table*}[t!]
\centering
\fontsize{7.5}{10}\selectfont
\caption{\fontsize{7.5}{10}\footnotesize Top-3 prediction results of GENRE*-\base, GENRE*-\basen, and GENRE* on NQ dev set in KILT. Highlights on the correct target sequence.} 
\begin{tabular}{ m{4.5cm} m{11cm}}
    \toprule
    \textbf{Query} & \textbf{Prediction Results} \\
    \midrule
        \multirow{4}{=}{\\[-2em]what do the 3 dots mean in math}
        & \textbf{GENRE*-\base} \hspace{5.5em} \colorbox{yellow}{Therefore sign}, Infinity symbol, Equation\\
        \cmidrule(lr){2-2}
        & \textbf{GENRE*-\basen} \hspace{1.5em} Slashed zero, Homo sapiens, Equation\\
        \cmidrule(lr){2-2}
        & \textbf{GENRE*} \hspace{6em} Ellipsis, Infinity symbol, Homo sapiens\\
    \midrule
        \multirow{6}{=}{\\[-4em]rizal finished all the chapters of the novel noli me tangere in} 
        & \textbf{GENRE*-\base}\hspace{0.85em} \colorbox{yellow}{Noli Me Tángere (novel)}, Noli Me Tangere (opera), Noli Me Tangere (Bernini)\\
        \cmidrule(lr){2-2}
        & \textbf{GENRE*-\basen}\hspace{0.85em} Noli me tangere, \colorbox{yellow}{Noli Me Tángere (novel)}, Noli Me Tangere (opera)\\
        \cmidrule(lr){2-2}
        & \textbf{GENRE*}\hspace{2em} Noli me tangere, Non è l'inferno, Noli Me Tangere (opera)\\
    \bottomrule
\end{tabular}
\label{table: sub_lex}
\end{table*}

\begin{table*}[t!]
\centering
\fontsize{7.5}{10}\selectfont
\caption
     {\fontsize{6.5}{10}\footnotesize R-precision(\%) for the test sets of document retrieval tasks on datasets in KILT. Both GENRE* and GENRE*-\base are trained with NQ + TQA; other datasets are not seen during the training time. Best in \bf{Bold}.}
    \begin{tabular}{ccc|ccccccccc|c}
    \toprule
    & \multicolumn{2}{c}{In-Domain Datasets} & \multicolumn{10}{c}{Out-of-Domain Datasets (Inference Only, Zero-Shot)} \\
    \cmidrule(lr){2-3} \cmidrule(lr){4-13}  
    &NQ & TQA & FEVER & AY2 &WnWi& WnCw  & T-REX & zsRE & HoPo & ELI5 & WoW & OoD Avg \\
    \midrule
    GENRE* & 52.7 & 64.8 & 64.2 & 9.1 & 2.8& 3.4 & 53.9 & 76.1 & 34.3 & 11.2 & 48.9 & 33.8 \\
    GENRE*-\base & \bf{59.4} & \bf{68.7} & \bf{67.0} & \bf{10.3} & \bf{5.4} & \bf{7.8} & \bf{59.1} & \bf{79.2} & \bf{37.5} & \bf{12.5} & \bf{51.7} & \bf{36.7}\\
    \bottomrule
    \end{tabular}
\label{table: zeroshot}
\end{table*}

\paragraph{(1) Coherency between Generative Retrieval and \emb}
Table~\ref{table: kilt} shows that \async, which replaces \mat using the encoder of \ret every $N$ epochs, tends to show higher performance than \Base, which uses fixed \mat. 
Also, updating \mat more frequently (smaller $N$) leads to better performance. 
Such results suggest that having high coherency between \ret and \emb improves the performance. 
However, as it needs extra cost to construct \mat for each update, there is a tradeoff between the computation overhead and the performance.

\paragraph{(2) Contrastive Learning}
Table~\ref{table: kilt},\ref{table: NQ320k},\ref{table: Hotpot} show that \pipe shows consistently higher performance than \base. 
The results indicate that \emb trained on contrastive loss tends to construct better \mat by leveraging the benefits of contrastive learning of constructing well-structured overall embedding space and regularizing the space to be uniform~\citep{ni2021sentence, Gao2021COILRE, Gao2021SimCSESC, Izacard2021UnsupervisedDI}.
When we calculate $L_\text{uniformity}$, a metric which checks how well the embedding space is constructed~\citep{Wang2020UnderstandingCR}, \pipe (-19.7) shows a lower number than \base (-18.2) where the lower the better.

\begin{figure}[t!]
\vspace{-1em}
\centering
\includegraphics[width=0.4\textwidth]{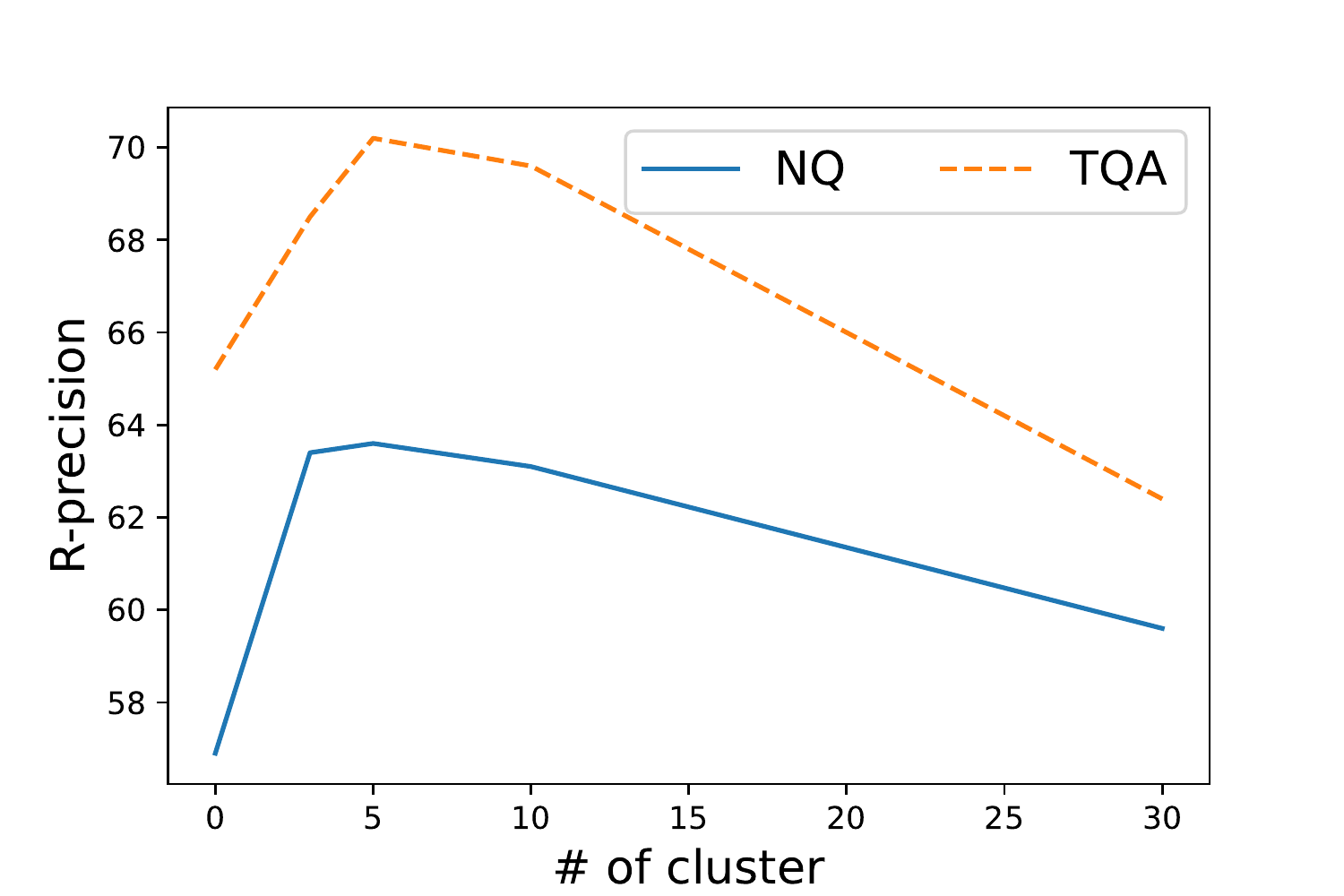}
\caption{\fontsize{7.5}{10}\footnotesize Effect of the maximum number of clusters (number of contextualized vocab embeddings) for each token on the performance of GENRE*-\base. The results when the number of clusters is zero are the results of GENRE*.}
\label{fig:cluster_num}
\end{figure}

\paragraph{(3) Contextualized Embeddings Size}
\mat contains multiple contextualized embeddings for each token where the number of embeddings per token is controlled by the clustering method.
In Figure~\ref{fig:cluster_num}, the performance shows the highest performance when using a maximum of five contextualized embeddings per token, and using a number higher or lower than five tends to decrease the performance for both NQ and TQA. 
This suggests that having too many vocab embeddings can be distracting while having too few can be not representative enough; the number of clusters should be neither too few nor too many. 

\vspace{-0.5em}
\paragraph{(4) Longer Context}
We compare the results between \basen and \base where \basen is a variant of \base where \mat is constructed using shorter context (\textit{only the title} without the document content) as input to \emb.
While both \base and \basen use \mat as decoder vocab embeddings, the contextualized vocab embeddings of \basen contain less contextual information compared to those of \base due to shorter context input to \emb.
Therefore, as shown in Table~\ref{table: sub_lex}, \basen performs poorly in cases where the document content is necessary for successful retrieval.
Additionally, \basen has lower R-precision in both NQ and TQA (58.4\% and 68.2\%) compared to \base (59.4\% and 68.7\%), indicating a correlation between performance and the amount of contextual information in \mat (non-parametric space). 


\section{Conclusion}
In this paper, we propose \Npd~(\npd), a new decoding method that can be applied to canonical generative retrieval models by simply replacing the decoder vocab embeddings from vanilla vocab embeddings to nonparametric contextualized vocab embeddings (output embeddings of an encoder).
This way, the generative retrieval does not rely solely on the information encoded in its own model parameters but can also utilize the information encoded in the contextualized embeddings. Using \npd in generative retrieval significantly improves the performance, achieves higher data and parameter efficiency, and shows more robustness in the zero-shot setting. In future work, we plan to apply \npd to various other tasks beyond task retrieval.


\section{Limitations}
\npd uses k-means clustering to reduce the number of contextualized embeddings, the performance varies by how the contextualized embeddings are clustered. As the process is relatively inconsistent, reducing the number with other methods would make the model performance more consistent.
Also, as it is not trivial to add new contextualized token embeddings on top of pre-constructed \mat due to the clustering step, we did not perform on dynamic corpus setup where new items are added or updated. 

\npd is applicable to all generative retrieval models including GMR or SEAL which needs all token embeddings, however, we focused on generative retrieval models with representative output as the retrieval target in this work. 
Also, while it is a general approach applicable to all encoder-decoder models, we focused on applying the method to T5.



\section*{Acknowledgements}
We thank Seonghyeon Ye, Joel Jang, and Sejune Joo for constructive feedback.
This work was partly supported by Kakao Brain grant (2021, Memory-Augmented Language Model, 80\%) and Institute of Information \& communications Technology Planning \& Evaluation (IITP) grant funded by the Korea government (MSIT) (No.2022-0-00264, Comprehensive Video Understanding and Generation with Knowledge-based Deep Logic Neural Network, 20\%).

\bibliography{custom}
\bibliographystyle{acl_natbib}

\appendix


\section{Nonparametric Decoding}
\label{app:overall_model}

\subsection{Different Types of Contrastive Loss for \Pipe} \label{app: pipe_details}
We experiment with three different types of contrastive loss when training \pipe. In this section, we show the losses and how the results differ by each loss.

Given a training dataset of pairs $\{(\textbf{\textit{q}}, \textbf{\textit{t}})\}$ where $\textbf{\textit{q}}$ is the query text, and $\textbf{\textit{t}}$ is the retrieval target (title of the document to retrieve) composed of multiple tokens $\textbf{\textit{t}}_{i}$ ($1 \le i \le k$ where $k$ is the length of the target), we split all tokens into $k$ separate pairs $\{(\textbf{\textit{q}}, \textbf{\textit{t}}_i)\}$ to construct the training dataset of query-token.
The three loss differs in what the model considers as a negative set and a positive set.

\paragraph{Loss 1: \textbf{Neg}: In-Batch Negatives / \textbf{Pos}: Single Token Embedding}
With the query-token dataset, we train \ret's first output token representation from the decoder to be close to all $\textbf{t}^{+} \in \{\textbf{t}_1, \cdots, \textbf{t}_k\}$ (embedding of any token in the retrieval target $\textbf{\textit{t}}$) given the query $\textbf{\textit{q}}$ as an input to \ret.
The objective is to minimize the contrastive loss to make the query text embedding  $\mathbf{q}$ be closer to positive token embedding $\mathbf{t}^{+}$:
\begin{gather} \label{eq}
L(\mathbf{q}, \mathbf{t}^{+}, \mathbf{t}_1^{-}, \cdots, \mathbf{t}_{|\mathcal{T}^-|}^{-}) \\ = -\log \frac{e^{<\mathbf{q}, \mathbf{t}^+>}}{e^{<\mathbf{q}, \mathbf{t}^+>} + \sum_{\mathbf{t}^- \in \mathcal{T}^-}{e^{<\mathbf{q}, \mathbf{t}^->}}}
\end{gather}
where $\langle  \; , \; \rangle$ is the inner product value between the two embeddings, and $\mathcal{T}^- = \{\mathbf{t}_1^{-}, \cdots, \mathbf{t}_{|\mathcal{T}^-|}^{-}\}$ is the set of negative token embeddings, which are other token embeddings in the training batch that are not paired with $\textbf{\textit{q}}$ (in-batch negatives~\citep{karpukhin2020dense}).

\paragraph{Loss 2: \textbf{Neg}: Contextualized Embedding Matrix / \textbf{Pos}: Single Token Embedding}
The loss differs from the upper loss in that it considers \textit{all} embeddings in contextualized embedding matrix except the single positive embedding as negative rather than performing the in-batch negatives which consider the subset of contextualized embedding matrix as negatives. 
The equation is same as Equation~\ref{eq}, but elements in $\mathcal{T}^-$ are \textit{all} other token embeddings in contextualized embedding matrix.

\paragraph{Loss 3: \textbf{Neg}: Contextualized Embedding Matrix / \textbf{Pos}: Multiple Token Embedding}
The loss differs from the upper loss in that it considers \textit{all} token embeddings in the title as positive embeddings; for each query $\textbf{\textit{q}}$, there are more than one positive contextualized token embeddings.

With the query-token dataset, where $\mathcal{T}^+ = \{\textbf{t}_1^+, \cdots, \textbf{t}_k^+\}$, set of positive token embeddings, we train \ret's first output token representation from the decoder to be close to all token embeddings in $\mathcal{T}^+$ given the query $\textbf{\textit{q}}$ as an input to \ret.
The objective is to minimize the contrastive loss to make the query text embedding  $\mathbf{q}$ be closer to all positive token embedding in $\mathcal{T}^+$:
\begin{gather}
L(\mathbf{q}, \mathbf{t}_1^{+}, \cdots, \mathbf{t}_{|\mathcal{T}^+|}^{+},\mathbf{t}_1^{-}, \cdots, \mathbf{t}_{|\mathcal{T}^-|}^{-}) \\ = -\log \frac{\sum_{\mathbf{t}^+ \in \mathcal{T}^+}{e^{<\mathbf{q}, \mathbf{t}^+>}}}{\sum_{\mathbf{t}^+ \in \mathcal{T}^+}{e^{<\mathbf{q}, \mathbf{t}^+>}} + \sum_{\mathbf{t}^- \in \mathcal{T}^-}{e^{<\mathbf{q}, \mathbf{t}^->}}}
\end{gather}
where $\langle  \; , \; \rangle$ is the inner product value between the two embeddings, and $\mathcal{T}^- = \{\mathbf{t}_1^{-}, \cdots, \mathbf{t}_{|\mathcal{T}^-|}^{-}\}$ is the set of negative token embeddings, which are \textit{all} other token embeddings in contextualized embedding matrix.

\subsection{Clustering} \label{app: clustering}
\paragraph{Example}
When a token ``the" appears in the corpus 100 times, 100 different contextualized embeddings of ``the'' are extracted by the encoder model at first. 
Then, we perform k-means clustering on the 100 contextualized embeddings to cluster them into at most $k$ clusters and save all centroid embeddings.
We leave only the $k$ centroid embeddings as the decoder vocab embeddings of the token ``the'' and assign a new decoder token ID for each contextualized embedding by the cluster it belongs to.
By repeating the process over all the tokens, each token has a number of contextualized embeddings less or equal to $k$. As there are multiple contextualized token embeddings for a single token, we replace the ground-truth target token IDs with the newly constructed decoder token IDs to specify which contextualized token embedding the ground-truth target token ID is referring to. 

\paragraph{Storage Footprint}
We analyzed how much storage footprint can be saved through the clustering method. When we use the KILT version Wikipedia corpus and titles as the retrieval target, about 37M token embeddings need to be stored. If the maximum number of token embeddings per token is set to 5 ($k=5$), only about 117K token embeddings need to be stored. Also, vanilla vocab embeddings of T5 ($k=1$) use about 32K token embeddings.
Therefore, when $k=5$, it only needs 0.3\% of the storage footprint compared to when storing all token embeddings of the title and about 3.7 times more storage compared to the vanilla vocab embeddings.
When $k=5$, it needs 0.47GB of storage footprint to save all the vocab embeddings, whereas the vanilla vocab embeddings ($k=1$) need 0.13GB. The increase in the storage footprint of vocab embeddings (0.34GB) is marginal compared to the storage footprint to save the model parameters (3GB).

\section{Experimental Setup} \label{app:exp_ret}
\subsection{GENRE* and GENRE* with \npd}
We train all models using a pre-trained T5-large~\citep{raffel2020exploring} checkpoint from \citet{Wolf2020TransformersSN} as the initial checkpoint (770M parameters). 
GENRE* and all generative retrieval models with \npd are trained with the same hyperparameter setting for a fair comparison. 
The training was done on 8 32GB V100 GPUs or a similar device. 
We train using Adafactor with a learning rate 1e-4\footnote{We also tried with a learning rate of 1e-3, a commonly used learning rate, but le-4 shows consistently higher performance.} with a linear warm-up for the first 10\% of training and then linear decay with batch size 512 till a maximum of 150 epochs with early stopping. All results are from a single run.

\subsection{BM25 \& DPR} \label{app: setup2}
To match the setting (dataset) similar to other baseline models, we train DPR~\citep{karpukhin2020dense} in a document retrieval task. Unlike \citet{Maillard2021MultiTaskRF}, which performs document retrieval tasks by training the model on passage-level tasks and considers the retrieval successful if it retrieves the passage in the target document, we train DPR on document-level tasks so that it retrieves the document itself.
We consider the first five paragraphs as the content and train the model so that the query embedding gets close to not the paragraph embedding but the document embedding. We use only the first five paragraphs of each document due to the limit in input length, and to keep it the same as the information used when dumping \mat by \emb.
The number of the corpus in the document retrieval tasks is the same as the number of pages in the KILT dataset. 
For BM25, we use pyserini~\citep{Lin_etal_SIGIR2021_Pyserini} where the corpus is the same as in DPR. All results are from a single run.

\subsection{Datasets}
For zero-shot evaluation, we also evaluated over WnWi~\citep{DBLP:conf/cikm/GuoB14} and ELI5~\citep{Fan2019ELI5LF}.


\subsection{Prefix Tree}\label{sec4:inference}
We perform a constrained beam search with prefix tree~\citep{cao2021autoregressive} during the inference step to assure that all generated sequences are in the corpus. The prefix tree is constructed with the tokenization result of the corpus, and we perform a constrained beam search by masking out the tokens that do not create a sub-string of the text in the corpus. We find the next tokens from the top-k of the unmasked ones. 
While \textit{token ID} was used as the node of the prefix tree in previous works since each token was mapped to a unique token ID, we construct a prefix tree with the \textit{text} of the token as the node, because \mat contains multiple token IDs for a single token. Therefore, rather than unmasking only a single token ID, we unmask all token IDs that correspond to the text in order to unmask a token. We keep the beam size to 10 for all experiments following \citet{cao2021autoregressive}.

\section{Experimental Results} \label{app:analysis}

\subsection{Multi-hop Dataset} \label{app: hotpot}
\paragraph{GENRE* in multi-hop setting}
During the inference step of GENRE*, in the first hop, GENRE* retrieves the title of the most relevant document (T1) when given a query, and in the second hop, GENRE* retrieves the title of the most relevant document (T2) when given the query, T1, and the context of T1 as input to the model.

\paragraph{Bridge vs. Comparison Questions}
HotpotQA contains both the bridge and the comparison questions; bridge questions are those that need to infer the missing intermediate entity from the document content of the first hop, and comparison questions are those with the two entities mentioned simultaneously.
We analyzed the performance of MDR, GMR, GENRE*, and GENRE* with \npd (GENRE*-\base, GENRE*-\pipe) by dividing the performance into bridge and comparison questions. 
GENRE* with \npd shows the highest performance in comparison questions among all models, and the highest performance in bridge questions among the generative retrieval models. 

\begin{table}[t!]
\centering
\fontsize{7.5}{10}\selectfont
\caption
     {\fontsize{6.5}{10}\footnotesize HotpotQA bridge vs. comparison (Top2)}
    \begin{tabular}{ccc}
    \toprule  
        & Bridge & Comparison \\
    \midrule
    MDR & \textbf{58.7} & 94.8 \\
    \midrule
    GMR & 47.9 & 96.4  \\
    GENRE* & 47.2 & 91.4 \\
    GENRE*-\base & 53.1 & 96.6 \\
    GENRE*-\pipe & 54.9& \textbf{96.9}\\
    \bottomrule
    \end{tabular}
\label{table: hotpot_analysis}
\end{table}

\subsection{Benefits of \Npd}

\begin{table*}[t!]
\centering
\fontsize{7.5}{10}\selectfont
    \begin{tabular}{ccc|cc|cccc|cc}
    \toprule
          Training Dataset & \multicolumn{2}{c}{Single ($\leq$ 3\%)}& \multicolumn{2}{c}{NQ+TQA ($\leq$ 5\%)}& \multicolumn{4}{c}{NQ+TQA+HotpotQA+ELI5 ($\leq$ 16\%)} & \multicolumn{2}{c}{All KILT (100\%)} \\      
    \midrule
    \midrule
         Model & NQ & TQA & NQ & TQA & NQ & TQA  & ELI5& HotpotQA & NQ & TQA \\
    \midrule
    GENRE & - & - & - & - & 58.3 &\textbf{69.6}&13.2&40.3& 60.3 & \textbf{69.2} \\
    GENRE* & 51.8 & 65.0 & 52.7 & 64.8 &54.2&67.8&13.8& 43.5&- & -  \\
    GENRE*-\base  & 59.0 & 68.2 & 59.4 & 68.7 &59.3&68.9& 14.2& 44.9&- \\
    GENRE*-\async& 59.2 & 68.4 & 59.8 & 68.7 &60.1&69.1&14.0& 46.3 & -&- \\
    GENRE*-\pipe& 59.8 & \textbf{68.6} & \textbf{60.3} & \textbf{68.9}&\textbf{60.7}&68.6&\textbf{14.9} & \textbf{47.0} & -&-\\
    \midrule 
    BM25 & ${23.4}^\dagger$ & ${25.2}^\dagger$ & ${23.4}^\dagger$ & ${25.2}^\dagger$ &${23.4}^\dagger$ & ${25.2}^\dagger$& ${5.3}^\dagger$ & ${38.4}^\dagger$& ${23.4}^\dagger$ & ${25.2}^\dagger$\\
    DPR & ${\textbf{60.1}}^\dagger$ & ${63.9}^\dagger$ & ${59.5}^\dagger$ & ${62.9}^\dagger$ &${58.0}^\dagger$&${63.2}^\dagger$&${10.6}^\dagger$&${39.3}^\dagger$& 59.4 & 61.5\\
    SEAL & - & - & - & -&-&-&-&- & \textbf{63.2} & 68.4 \\
    \bottomrule
    \end{tabular}
\caption
     {\fontsize{6.5}{10}\footnotesize R-precision(\%) for document retrieval task on NQ and TQA test dataset (KILT version). Results except for GENRE* and GENRE* with \npd (GENRE*-\base, GENRE*-\async, and GENRE*-\pipe) are from the KILT leaderboard. 
     The column of the table is divided by how many training datasets are used. 
     Numbers in the bracket are the rate of the number of training datasets over the number of training datasets when using all KILT datasets. 
     Results of GENRE and SEAL are from \citet{cao2021autoregressive} and \citet{Bevilacqua2022}, respectively. Results with $\dagger$ in BM25 and DPR are trained in the same setting as GENRE* with \npd (Appendix~\ref{app: setup2}). Best in \bf{bold}.}
\label{table: kilt_multi}
\end{table*}

\paragraph{Multitask Training} \label{app: kilt-multi}
Results in Table~\ref{table: kilt_multi} show that GENRE*-\pipe outperforms GENRE* by 6\% in single-task which demonstrates the effectiveness of \npd.
For both cases where the model is trained over a single dataset and over NQ and TQA together (NQ+TQA), GENRE* with \npd shows higher performance over GENRE*.
Note that due to limited available resources, we did not train GENRE* with \npd on the full KILT dataset (ALL KILT) as in GENRE\footnote{GENRE uses 128 V100 GPUs with 32GB of memory for about 33 hours.}, DPR, or SEAL. 
However, \pipe trained on less than 5\% of the training dataset from the full KILT dataset shows higher or comparable performance to those models. 

\begin{table*}[t!]
\centering
\fontsize{7.5}{10}\selectfont
\caption{\fontsize{7.5}{10}\footnotesize Top-3 prediction result of \base, and GENRE*} 
\begin{tabular}{ m{5cm} m{10cm}} 
    \toprule
    \textbf{Query} & \textbf{Prediction Result} \\
    \midrule
        \multirow{4}{=}{\\[-3em]what do the 3 dots mean in math}
        & \textbf{\base} \hspace{1em} \colorbox{yellow}{Therefore sign}, Infinity symbol, Equation\\
        \cmidrule(lr){2-2}
        & \textbf{GENRE} \hspace{2em} Ellipsis, Infinity symbol, Homo sapiens\\
    \midrule
        \multirow{4}{=}{\\[-4em]what does the pearl symbolize in the bible} 
        & \textbf{\base} \hspace{2em} \colorbox{yellow}{Parable of the Pearl}, Mitzvah, Pearl of Wisdom\\
        \cmidrule(lr){2-2}
        & \textbf{GENRE}\hspace{2em}  Pearl of Great Price, Perlin, Promised Land \\
    \midrule
        \multirow{4}{=}{\\[-3em]does archie end up with betty or veronica in riverdale} 
        & \textbf{\base}\hspace{2em}  \colorbox{yellow}{Archie Marries Veronica/Archies Marries Betty}, List of Riverdale characters, Archie Buchanan \\
        \cmidrule(lr){2-2}
        & \textbf{GENRE} \hspace{2em} Riverdale (2017 TV series), List of Riverdale characters, Archie Mitchell\\
    \midrule
        \multirow{4}{=}{\\[-4em]actor who plays dr avery on grey's anatomy} 
        & \textbf{\base} \hspace{2em}\colorbox{yellow}{Jesse Williams (actor)}, Jesse Williams, Jesse Spencer\\
        \cmidrule(lr){2-2}
        & \textbf{GENRE}\hspace{2em} Marc Alaimo, Patrick Warburton, Jeffrey Dean Morgan\\
    \midrule
        \multirow{4}{=}{\\[-3em]when did equus first appear in fossil record}
        & \textbf{\base}\hspace{1.4em} \colorbox{yellow}{Evolution of the horse}, Equis, Eurydice \\
        \cmidrule(lr){2-2}
        & \textbf{GENRE}\hspace{2em}Equidae, Equis, Equinox\\
    \midrule
        \multirow{4}{=}{\\[-2em]who decides the number of judges in the high court}
        & \textbf{\base}\hspace{2em} \colorbox{yellow}{Indian High Courts Act 1861}, High Court of Australia, Supreme Court of India\\
        \cmidrule(lr){2-2}
        & \textbf{GENRE}\hspace{2em}Supreme Court of the United Kingdom, Supreme Court of India, High Court of Australia\\
    \midrule
        \multirow{4}{=}{\\[-3em]when's the last time the philadelphia eagles played the new england patriots}
        & \textbf{\base}  \colorbox{yellow}{Super Bowl XXXIX}, New England Patriots, Super Bowl XXXVIII\\
        \cmidrule(lr){2-2}
        & \textbf{GENRE}\hspace{2em}New England Patriots, Philadelphia Eagles, History of the Philadelphia Eagles\\
    \midrule
        \multirow{4}{=}{\\[-2.7em]rizal finished all the chapters of the novel noli me tangere in} 
        & \textbf{\base}\hspace{0.85em} \colorbox{yellow}{Noli Me Tángere (novel)}, Noli Me Tangere (opera), Noli Me Tangere (Bernini)\\
        \cmidrule(lr){2-2}
        & \textbf{GENRE}\hspace{2em} Noli me tangere, Non è l'inferno, Noli Me Tangere (opera)\\
    \midrule
        \multirow{6}{=}{\\[-6em]during which season does cape town receive rainfall} 
        & \textbf{\base}\hspace{1.4em} \colorbox{yellow}{Climate of South Africa}, City of Cape Town, Cape Town water crisis\\
        \cmidrule(lr){2-2}
        & \textbf{GENRE}\hspace{2em} Cape Town, City of Cape Town, Cape Town water crisis\\
    \bottomrule
\end{tabular}
\label{table: analysis_lex_prediction}
\end{table*}


\paragraph{Robust to Low Lexical Overlap} \label{app:lexical}

We first run TF-IDF over all the queries of the NQ dev set in KILT and divide the queries into two sets: low-overlap\footnote{e.g., Q: During which season does cape town receive rainfall / Target Document: Climate of South Africa} and high-overlap\footnote{e.g., Q: where was the \textit{world economic forum} held this year / Target Document: World Economic Forum}. Low-overlap is a set of queries with a TF-IDF score lower than average, and high-overlap is the rest of the queries.

Generative retrieval with \npd shows especially strong performance on queries in the low-overlap set; queries that in most cases need the context information unless the model saw the information during the training step.
We check four sets:\\
1. GENRE+/\base+: queries where both \base and GENRE* successfully retrieved\\
2. GENRE+/\base-: queries where GENRE* successfully retrieved and \base failed\\
3. GENRE-/\base+: queries where GENRE* failed and \base succeed \\
4. GENRE-/\base-: queries where GENRE* and \base both failed.

\begin{figure}
    \centering
    \begin{minipage}[b]{0.5\textwidth}
    \centering
    \includegraphics[width=0.8\textwidth]{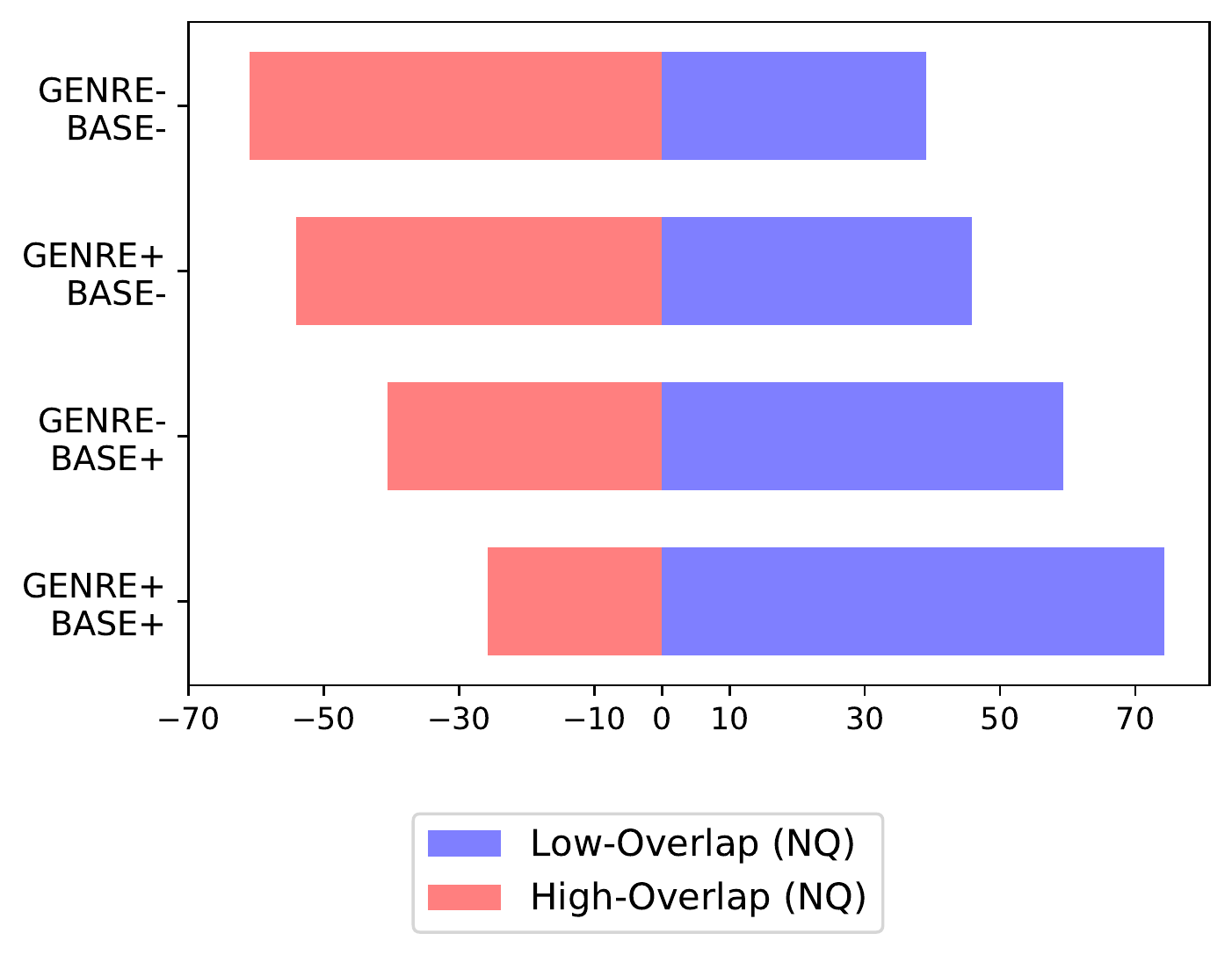}
    \caption{Red bar indicates the high rate and the blue bar indicates the low rate. The rate is measured by NQ dev set in KILT. Details about high-rate and low-rate is in Appendix~\ref{app:lexical}.}
    \label{fig:nq_compare_lexical}
    \end{minipage}
    \begin{minipage}[b]{0.5\textwidth}
    \centering
    \includegraphics[width=0.8\textwidth]{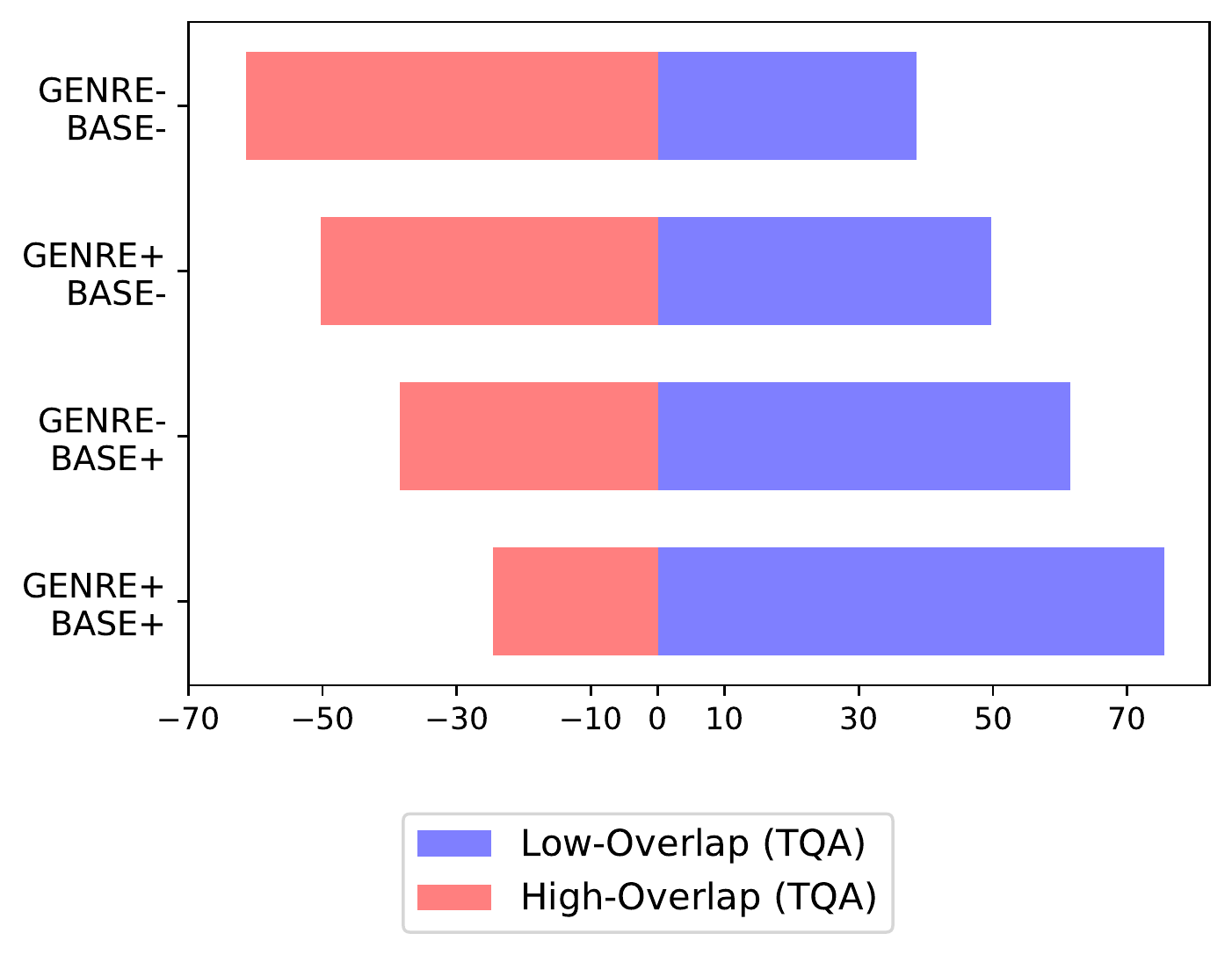}
    \caption{Red bar indicates the high rate and the blue bar indicates the low rate. The rate is measured by TQA dev set in KILT. Details about high-rate and low-rate is in Appendix~\ref{app:lexical}.}
    \label{fig:tqa_compare_lexical}
    \end{minipage}
\end{figure}

Figure~\ref{fig:nq_compare_lexical} and Figure~\ref{fig:tqa_compare_lexical} show the low-rate (blue) and high-rate (red).
Low-rate of each case is calculated as $\frac{\{Q \cap L\}}{Q}$, where $Q$ is a set of queries in each case and $L$ is a set of queries in a low-overlap set.
High-rate of each case is calculated as
$\frac{\{Q \cap H\}}{Q}$, where $H$ is a set of queries in a high-overlap set.

For both figures, GENRE-/\base+ shows a higher number in low-rate, which indicates that \base tend to successfully predict queries in the low-overlap set compared to GENRE*.
Also, for both figures, GENRE+/\base+ shows a high number in low-rate and GENRE-/\base- shows a high number of high-rate, which indicates that queries in the high-overlap set tend to be easy questions for both GENRE and \base whereas queries in the low-overlap set are difficult for both models.

Also, Table~\ref{table: analysis_lex_prediction} shows samples of the top-5 prediction results of \base and GENRE* where \base successfully retrieved the correct item and GENRE* failed. 
Moreover, Table~\ref{table: overlap} shows the performance of GENRE and GENRE*-\base for low-overlap and high-overlap sets.
The results suggest that \base is robust on queries in the low-overlap set compared to GENRE*.

\begin{table}[t!]
\centering
\fontsize{7.5}{10}\selectfont
\caption
     {\fontsize{7.5}{10}\footnotesize R-precision(\%) for the document retrieval task on NQ dev dataset in KILT. See details about Low and High Overlap in Appendix~\ref{app:lexical}.}
    \begin{tabular}{cccc}
    \toprule
    & GENRE* & GENRE*-\basen & GENRE*-\base\\
    \midrule
    Low-Overlap & 45.8 & 51.6 & 52.7 \\
    High-Overlap & 71.3 & 75.3 & 75.8 \\
    \midrule
    Total & 58.3 & 63.2 & 64.0 \\
    \bottomrule
    \end{tabular}
\label{table: overlap}
\end{table}

\begin{table}[t!]
\centering
\fontsize{7.5}{10}\selectfont
\caption
     {\fontsize{6.5}{10}\footnotesize R-precision(\%) for the document retrieval task on NQ and TQA test dataset in KILT.
     We compare the results of GENRE*, \basen, and \base where the models are trained with NQ+TQA. The results show the importance of extracting contextualized embeddings with not only the title but also the corresponding document content.}
    \begin{tabular}{cccccccc}
    \toprule  
    & GENRE* & GENRE*-\basen & GENRE*-\base \\
    \midrule
    NQ & 52.7 & 58.4 & \textbf{59.4} \\
    Trivia & 64.8 & 68.2 & \textbf{68.7}  \\
    \bottomrule
    \end{tabular}
\label{table: add_context}
\end{table}

\subsection{What is Well-Constructed Contextualized Embedding Matrix~(\mat)?} \label{app: ce}
In this section, we analyze \npd with GENRE* so we skip the model name.

\paragraph{(1) Having High Coherency between Generative Retrieval and \emb} \label{ablation: freq}
We analyzed how the performance changes according to how often the replacement of \emb by the encoder of the generative retrieval occurs (replacement for every $N$ epoch) with \asyncn.
When comparing the performance with $N=\{10, 20, 50\}$, \asyncn shows the highest performance at $N=10$, and the performance tends to deteriorate as N becomes larger. 
Also, all \asyncn show higher performance than \basen, which uses \mat with no replacement during training ($N$ = max training epoch).
Results show that although the model requires high computation cost and longer training time as $N$ gets smaller, it is important to have high coherency between the contextualized embeddings (output embeddings of \emb) and \ret by frequent replacement. 

\begin{table*}[t!]
\centering
\fontsize{7.5}{10}\selectfont
\caption
     {\fontsize{7.5}{10}\footnotesize R-precision(\%) for the document retrieval task on NQ and TQA test dataset in KILT. See Appendix~\ref{ablation: contrastive} for details about how the loss term differs. The loss term is used while training \pipe in contrastive learning (step 1 of training \pipe).}
    \begin{tabular}{cc|cc}
    \toprule
    Positive & Negative & NQ & TQA\\
    \midrule
    Single Token Emb & In-Batch Negatives & 60.0 &  \bf{68.9}\\
    Single Token Emb & Contextualized Embedding Matrix &58.9 & 68.4\\
    Multiple Token Emb & Contextualized Embedding Matrix  &\bf{60.3} & \bf{68.9} \\
    \bottomrule
    \end{tabular}
\label{table: contra}
\end{table*}

\paragraph{(2) Training \mat with Contrastive Learning} \label{ablation: contrastive}
Appendix~\ref{app: pipe_details} shows the details of three different contrastive losses that we experiment over when training \Pipe. 
Table~\ref{table: contra} show the performance of \pipe with different contrastive loss, which differs by what is considered as the positive pair and the negative pair. 
\textit{Multiple Token Emb} considers all token embeddings in the same target sequence as positive pairs, and \textit{Single Token Emb} considers all token embeddings separately thus only one of the token embedding from the title token embeddings is considered as positive pair.
\textit{In-Batch Negatives} considers all embeddings in a batch except for the positive embedding as negative pairs, and \textit{Contextualized Embedding Matrix} considers all embeddings in the contextualized embedding matrix (a matrix constructed with the contextualized token embeddings) except for the positive embeddings as negative pairs.

The model trained on contrastive loss with multiple token embeddings as positive pairs, and all other embeddings in contextualized embedding matrix as negative pairs (Loss3) show the highest performance. 
The model trained on the same negative but with a single token embedding as positive (Loss2) shows the lowest performance.
The model with single token embedding as positive and in-batch negatives as negative pairs (Loss1) shows the performance in-between.

As in \citet{Xiong2021ApproximateNN}, the model with Loss2 and Loss3 has the benefits of looking at the global embedding space by considering the contextualized embedding matrix as the negative pair, unlike Loss1 which only considers embeddings in the same batch as negatives (in-batch negatives).
However, Loss2 show lower performance than Loss1 as in the case where the model considers a single token embedding as a positive pair, the model considers the rest of the token embeddings in the same title as the negative pair. As the token embeddings in the same title are matched with the same query, such a training method seems to make the model confused and leads to bad performance. 
Thus when considering a single token embedding as positive pair (Loss1 or Loss2), it is better to consider only the embeddings in the same batch as negatives (in-batch negatives) rather than on all the token embeddings (Contextualized Embedding Matrix) as there is a low possibility of the model to have two different token embeddings of the same title in a batch.

\paragraph{(3) Contextualized Embeddings Size} \label{ablation:cluster}
As saving all contextualized token embeddings to use as the vocab embedding matrix requires a large storage footprint ($\approx$ 148GB), we reduce the number of token embeddings by clustering and saving only the $k$ centroid embeddings for each token (Section~\ref{sec4: cluster}).
Figure~\ref{fig:cluster_num} shows the effect of the maximum number of clusters for each token ($k$) on the performance. 
Models with a $k=5$ (maximum of five different contextualized token embeddings for each token) show the highest performance and having $k$ smaller or larger than five decreases the performance. 
We hypothesize that the performance of models with $k<5$ degrades because the number of the embeddings is too small to contain all different contextual meanings of the token and thus will be closer to vanilla token embedding.
In contrast, the performance of models with $k>5$ decreases because the search space of each generation step is too large and the parametric space of the model becomes too fine-grained which might distract the model.

\paragraph{(4) Longer Context} \label{app: oursn}

To see how informative the document context (length of the context) affects performance, we compared the performance of \base, \basen, and GENRE*.
GENRE*, which uses vanilla vocab embedding as the target embedding, has to depend solely on the information encoded in its own parameters (the parametric space of the generative retrieval model).
On the other hand, \base and \basen can depend on not only the parametric space of the generative retrieval model but also the non-parametric space of corpus information embedded in the contextualized target embedding. 
By utilizing the contextualized target embedding, the model can know in which context the token is used and discern documents with different contexts. 

Although both \base and \basen utilize contextualized target embeddings, the contextualized target embedding of \basen contains shorter context information compared to that of \base.
Therefore, \basen fails in cases where the document content is necessary to retrieve the target sequence successfully.
It is difficult for the model to predict the target without the help of the document content about what information is in the document or what relationship exists between the query and the target sequence. 
We can see from the table (Table~\ref{table: sub_lex}) that \base successfully retrieves as such information is embedded in the contextualized target embeddings whereas \basen fails as it does not contain the document content in its embeddings.
Also, Table~\ref{table: add_context} shows that there is a correlation between the performance and how much contextual information is encoded in the nonparametric space.

\paragraph{Characteristics by different \emb}  \label{analysis: analysis_emb}
We compare the contextualized token embeddings of \base, \async, and \pipe\footnote{We analyze the \emb of step2 in \pipe and last replace \emb for \async.}
For 1000 cluster embeddings, we check the rate of the same token among the top-5 embeddings similar to the corresponding embedding.
\base shows the lowest rate of 50\%. \async and \pipe show a similarly high rate of 70\%. 
The rate tends to increase as $N$ increases in \async.
Such results suggest that as the same token has a similar lexical meaning, it is better to have a relatively similar meaning. However, as the performance increases as a single token are matched to multiple token embeddings till $k=5$, it is also important to have slightly different meanings depending on the surrounding context. 
When checking which corpus bundles are bound to the same cluster, all three tend to depend on which \textit{position} of text the token is placed on and the \textit{meaning} of surrounding tokens.
For example, when we cluster a token ``Bee" into five clusters, it tends to group in: (1) word related to a person's name where ``Bee" appears in the middle of the name (Edmund Beecher Wilson), (2) word related to food where ``Bee" appears at the front of the word (Beef Jerkey), (3) ``Bee" with Spelling Bee (The 25th Annual Putnam County Spelling Bee) or the insect bee (Honey to the Bee) that appears at the end of a word, (4) word related to music where ``Bee" appears at the front of the word (Honey to the Bee), (5) word related to film or TV series where ``Bee" appears near the end (Queen Bees (TV series)). Such tendencies are shown in all three models. 

\begin{table*}[t!]
\centering
\fontsize{7.5}{10}\selectfont
\caption{\fontsize{7.5}{10}\footnotesize Examples of clusters when clustering over total contextualized token embeddings when \emb is the encoder of T5-large.} 
\begin{tabular}{m{1cm} m{1cm} m{11cm}} 
    \toprule
    \textbf{Cluster} & \textbf{Token} & \textbf{Documents} \\
    \midrule
        \textbf{1} &
        \textbf{\_Lincoln} &
        Moulton, Lincolnshire / Belton, North Lincolnshire / Walcott, Lincolnshire / Wrangle, Lincolnshire / Swineshead, Lincolnshire / Leverton, Lincolnshire / Kirton, Lincolnshire / Benington, Lincolnshire / Bicker, Lincolnshire / Dyke, Lincolnshire / Hilldyke, Lincolnshire / Waltham, Lincolnshire / Reepham, Lincolnshire / Bradley, Lincolnshire / Allington, Lincolnshire / Donington, Lincolnshire  ettleton, Lincolnshire / Panton, Lincolnshire / Beckingham, Lincolnshire / Bigby, Lincolnshire / ...\\
    \midrule
        \textbf{2} &
        \textbf{\_Squad} &
        Field hockey at the 2000 Summer Olympics – Men's team squads / Field hockey at the 2004 Summer Olympics – Men's team squads / Field hockey at the 1996 Summer Olympics – Men's team squads / Football at the 2000 Summer Olympics – Men's team squads / Football at the 1996 Summer Olympics – Men's team squads / List of Queensland rugby league team squads / Football at the 2006 Lusophony Games – Men's team squads / List of current AFL team squads / Football at the 1912 Summer Olympics – Men's team squads / List of New South Wales rugby league team squads / Football at the 1996 Summer Olympics – Women's team squads / Football at the 1988 Summer Olympics – Men's team squads / Football at the 1984 Summer Olympics – Men's team squads / Football at the 1976 Summer Olympics – Men's team squads / Football at the 1900 Summer Olympics – Men's team squads / Football at the 1904 Summer Olympics – Men's team squads / Football at the 1908 Summer Olympics – Men's team squads / Football at the 1992 Summer Olympics – Men's team squads / Football at the 1980 Summer Olympics – Men's team squads / Football at the 1972 Summer Olympics – Men's team squads / ...\\
    \midrule
        \multirow{9}{=}{\textbf{3}} &
        \textbf{\_Lincoln} & 
        William Lincoln Garver / Albert Lincoln Washburn / Charles Lincoln Edwards / Thomas Lincoln Casey Sr. / James Lincoln Collier / Abraham Lincoln Lewis / Earl Lincoln Poole / George Lincoln Goodale / George Lincoln Burr / Abraham Lincoln Keister / Elmer Lincoln Irey / Walter Lincoln Hawkins / Abram Lincoln Harris / Abraham Lincoln DeMond / Thomas Lincoln Tally / Abraham Lincoln Filene / Mary Lincoln Beckwith / Frederick Lincoln Emory / Howard Lincoln Hodgkins / Oliver Lincoln Lundquist / ...\\
        \cmidrule(lr){2-3}
        & \textbf{\textbf{\_Levi}} &
        John Levi Marti / John Levi Sheppard / Moses Levi Ehrenreich / Nathaniel Levi Gaines / Harry Levi Hollingworth / Thomas Levi Whittle / Austin Levi Fraser / George Levi Crane / Olin Levi Warner
        \\
        \cmidrule(lr){2-3}
        & \textbf{\textbf{\_Luke}} &
        Milledge Luke Bonham / Henry Luke Orombi / Henry Luke White / Vincent Luke Palmisano / George Luke Smith / Henry Luke Bolley / Mary Luke Tobin / James Luke Prendergast / John Luke Lowther / Jerry Luke LeBlanc / Thomas Luke Msusa / Robert Luke Deakin / Joseph Luke Cecchini\\
        \cmidrule(lr){2-3}
        & \textbf{\textbf{\_Lane}} &
        Carroll Lane Fenton\\
        \cmidrule(lr){2-3}
        & \textbf{\textbf{...}} & \\
    \midrule
        \multirow{8}{=}{\textbf{4}} &
        \textbf{\_Squad} & 
        True Story (Terror Squad album) / The Album (Terror Squad album)
        \\
        \cmidrule(lr){2-3}
        & \textbf{\textbf{\_Angel}} &
        Covenant (Morbid Angel album) / Domination (Morbid Angel album) / The Art of Dying (Death Angel album) / Act III (Death Angel album) / Heretic (Morbid Angel album)
        \\
        \cmidrule(lr){2-3}
        & \textbf{\textbf{\_Butterfly}} &
        Heavy (Iron Butterfly album) / Metamorphosis (Iron Butterfly album) / Ball (Iron Butterfly album)
        \\
        \cmidrule(lr){2-3}
        & \textbf{\textbf{\_Flip}} &
        Flip-flop (electronics) / Flipper (anatomy) / Respect Me (Lil' Flip album) / The Leprechaun (Lil' Flip album)
        \\
        \cmidrule(lr){2-3}
        & \textbf{\textbf{...}} & \\
    \bottomrule
\end{tabular}
\label{table: analysis_total_clustering}
\end{table*}

\paragraph{Clustering over total embeddings} \label{app:analysis_total}
To understand the spatial properties of the contextualized embeddings, we conducted a qualitative analysis on the embeddings, by performing k-means clustering over the total contextualized token embeddings of \base (\emb is the encoder of T5-large). Specifically, we clustered 36 million token embeddings, obtained from \emb, into 117,508 clusters\footnote{The number of the clusters is same as the number of the tokens in contextualized embedding matrix, hence same as the number of the clusters we used in ~\ref{sec4: cluster}.} using the FAISS k-means library~\citep{Johnson2021BillionScaleSS}. 

First, we randomly sampled 100 tokens, and for each token, we calculated the portion of the contextualized embeddings that belong to the top 10\% of the clusters which contain the most embeddings of the token. As a result, on average 67.6\% of the embeddings of a token are contained in the 10\% of the clusters which contain the token, with a standard deviation of 22.7. This indicates that most of the tokens are concentrated in a few spatial regions, while the others are spread over many different areas.

To get a deeper insight into the spatial properties of the embeddings, we picked two tokens, ``Lincoln" and ``Squad"  and visualized some of the clusters that contain the tokens(Table~\ref{table: analysis_total_clustering}). For each cluster, the tokens belonging to the cluster and their corresponding document names are shown. In (Table~\ref{table: analysis_total_clustering}), at most 20 documents are shown for each token and only 4 tokens are shown in cluster 3 and 4 for simplicity. The first and second examples show the case that a cluster is composed of only a single token, as mentioned above. Interestingly, all of the corresponding documents of the first cluster are related to Lincolnshire, a county of England. Similarly, the tokens in the second cluster are related to the documents about sports (usually football) squads. On the other hand, the third and fourth examples show the other case that a cluster contains only a few tokens that we are interested in. The members of the third cluster are related to the middle names, and a few embeddings of the token ``Lincoln" is contained in this cluster since there are some Wikipedia documents of the people whose middle name is Lincoln. Likewise, the fourth cluster consists of the embeddings which are related to the name of music albums(usually hip-hop and rock), where some of them are produced by the group named ``Blazin' squad", for example. These examples show how expressive can the contextualized embeddings be compared to the vanilla token embeddings; in this case, it is hard to expect that this various context-dependent information of a token can be sufficiently encoded into a single token embedding.

In summary, the results show that the contextualized embeddings corresponding to the same token are mapped to many different regions of the embedding space, depending on its context. This implies that the contextualized embeddings successfully acquired the contextual information of the corresponding documents, highlighting the effectiveness of utilizing contextualized embeddings for generative retrieval.


\end{document}